\documentclass[submitting]{nst}

\usepackage{amsmath,amssymb,bm,graphicx,booktabs,microtype,xcolor}
\usepackage{subfigure,dcolumn}
\usepackage[T2A,T1]{fontenc}
\usepackage[russian,english]{babel}

\usepackage{listings}
\newcommand{\pt}{p_{\mathrm{T}}}
\newcommand{\snn}{\sqrt{s_{\mathrm{NN}}}}

\newcommand{\Atwo}{A_{2}}
\newcommand{\NplusN}{\mbox{$N+N$}}
\newcommand{\NplusS}{\mbox{$N+S$}}
\newcommand{\Phirp}{\Phi_{\mathrm{RP}}}
\newcommand{\xperp}{\bm{x}_{\perp}}
\newcommand{\pperp}{\bm{p}_{\perp}}
\newcommand{\bvec}{\bm{b}}
\newcommand{\rvec}{\bm{r}}

\begin{document}

\title{Dilepton azimuthal anisotropy from coherent $J/\psi$ photoproduction in hadronic Pb--Pb collisions}

\author{Xueli Hua}
\affiliation{State Key Laboratory of Nuclear Physics and Technology, Institute of Quantum Matter, South China Normal University, Guangzhou 510006, China}
\affiliation{Key Laboratory of Atomic and Subatomic Structure and Quantum Control (MOE), Guangdong-Hong Kong Joint Laboratory of Quantum Matter, Guangzhou 510006, China}
\affiliation{Guangdong Basic Research Center of Excellence for Structure and Fundamental Interactions of Matter, Guangdong Provincial Key Laboratory of Nuclear Science, Guangzhou 
510006, China}

\author{Xin Wu}
\email[Corresponding author, ]{wuxinust@ustc.edu.cn}
\affiliation{State Key Laboratory of Particle Detection and Electronics, University of Science and Technology of China, Hefei 230026, China}

\author{Shuai Yang}
\email[Corresponding author, ]{syang@scnu.edu.cn}
\affiliation{State Key Laboratory of Nuclear Physics and Technology, Institute of Quantum Matter, South China Normal University, Guangzhou 510006, China}
\affiliation{Guangdong Basic Research Center of Excellence for Structure and Fundamental Interactions of Matter, Guangdong Provincial Key Laboratory of Nuclear Science, Guangzhou 
510006, China}

\author{Wangmei Zha}
\affiliation{State Key Laboratory of Particle Detection and Electronics, University of Science and Technology of China, Hefei 230026, China}

\begin{abstract}
Hadron-based event-plane estimators can be affected by nonflow and autocorrelation biases, especially at low multiplicity. Coherent $J/\psi$ photoproduction offers a complementary geometry reference through the correlation between quasireal-photon polarization and the reaction plane. We calculate the second-order dilepton coefficient $\Atwo$ relative to the reaction plane in hadronic Pb--Pb collisions at $\snn=5.36$~TeV within an equivalent-photon and generalized vector-meson-dominance Glauber framework. Nuclear diffraction and two-source interference produce a pronounced transverse-momentum dependence, while the rapidity dependence is weak. We also compare two treatments of coherent scattering in the overlap region of the two colliding nuclei. Suppressing the contribution from this region reduces the yield and strengthens the integrated polarization--reaction-plane correlation, with the largest differences in central and semicentral collisions and smaller differences toward peripheral collisions. Joint measurements of the coherent yield and $\Atwo$ can constrain how hadronic interactions disrupt nuclear coherence and clarify the photoproduction mechanism. With sufficient tagged yields, azimuthal correlations involving decay leptons from photoproduced $J/\psi$ mesons could complement studies of collectivity, with potential future applications to small systems.
\end{abstract}


\maketitle
\nolinenumbers

\section{Introduction}

Azimuthal anisotropies in heavy-ion collisions probe the quark--gluon plasma (QGP)~\cite{Adams:2005dq,Adcox:2004mh} by linking the initial spatial geometry to the medium's collective response and transport properties~\cite{Heinz:2013}. Extracting this information requires a reference for the collision orientation. Since the reaction plane is not directly observable, experiments rely on hadron-based estimators affected by finite multiplicity, fluctuations, and nonflow~\cite{Poskanzer:1998,Heinz:2013}. Their performance depends on multiplicity and rapidity coverage~\cite{Liu:2023,STAR:2022planes,Bilandzic:2010jr}. At low multiplicity, few-particle correlations can compete with collective signals, making an independent geometry reference particularly valuable. Recent O--O and Ne--Ne flow measurements~\cite{CMS:2026OO} extend this motivation to light-ion systems and to the question of whether $pA$ and high-multiplicity $pp$ collisions form a ``little drop'' of QGP~\cite{GrosseOetringhaus:2024}.

One way to access this geometry is through the strong electromagnetic fields generated by relativistic nuclei~\cite{Huang:2016}. In the equivalent-photon approximation, these fields are represented as fluxes of quasireal photons~\cite{Krauss:1997vr,Baur:2002,Baltz:2008}. Neutron tagging from nuclear Coulomb excitation has been proposed to select different impact-parameter distributions in exclusive photoproduction at electron--ion colliders~\cite{Wu:2025NST}. In nucleus--nucleus collisions, Breit--Wheeler dilepton transverse-momentum distributions are sensitive to the electromagnetic-field profile~\cite{Zha:2020BW,Brandenburg:2021}, while their azimuthal correlations reflect photon linear polarization~\cite{Li:2019,Xiao:2020,STAR:2021} and can constrain the initial geometry~\cite{Luo:2026NST}.

Photon polarization also carries information on the impact-parameter direction~\cite{Wu:2022}. The photons are linearly polarized along the locally radial transverse electric field. The polarization directions sampled across the target nucleus can therefore correlate with the impact-parameter vector. Because this reference does not use the soft hadrons in a subsequent bulk measurement, it can reduce hadronic nonflow and autocorrelation biases.

Coherent vector-meson photoproduction allows this polarization information to be read out through meson decays. The produced meson's linear polarization appears as a second-order azimuthal modulation in the angular distribution of its decay products~\cite{Criegee:1968,Ballam:1970,Schilling:1970,Mathieu:2018,ALICE:2023svb}. The decay angles therefore provide a measurable link to the impact-parameter direction. The underlying production process can be described in the vector-meson-dominance (VMD) picture, where a quasireal photon interacts with the target nucleus through its vector-meson components~\cite{Bauer:1977iq}. Combined with a Glauber treatment of nuclear multiple scattering, VMD provides a successful baseline for this process~\cite{Klein:1999qj,Miller:2007ri}. In symmetric nucleus--nucleus collisions, either nucleus can emit the photon, so the two indistinguishable emitter--target amplitudes interfere~\cite{Klein:2000,Brandenburg:2025review}.

Among vector mesons, the $J/\psi$ is of particular interest because its yield probes the hot medium through color screening, dissociation, and regeneration~\cite{Matsui:1986dk,Andronic:2015wma}. In collisions with hadronic overlap, this yield also receives a coherent-photoproduction contribution at very low $\pt$~\cite{ALICE:2015mzu,STAR:2019yox,LHCb:2021hoq,ALICE:2022zso,ALICE:2024whv}. Understanding this contribution is important for interpreting $J/\psi$ yields as medium probes~\cite{Shi:2017qep,Liu:2024photoproduction}.

Beyond yield measurements, STAR reported evidence for spin interference in ultraperipheral Au--Au and isobar collisions using the angle between the $J/\psi$ transverse momentum and the relative transverse momentum of its decay electrons~\cite{STAR:2026Jpsi}. In hadronic collisions, the accompanying particles also allow the decay angles to be studied relative to a reconstructed event plane. STAR results show a photon-induced $J/\psi$ decay anisotropy relative to the hadronic event plane in isobar collisions~\cite{Wang:2026STAR}, while CMS observed a negative second-order coefficient for coherent $J/\psi\to\mu^+\mu^-$ relative to the second-order event plane in peripheral Pb--Pb collisions at $\snn=5.36$~TeV~\cite{CMS:2026}. A color-dipole calculation describes the STAR isobar results and predicts the corresponding correlation at LHC energies~\cite{Shao:2025}.

Because the local photon-polarization direction varies across the transverse plane, changing the coherent source can change the decay anisotropy. In hadronic collisions, it remains uncertain whether the coherent source involves the full nuclei, mainly their spectator regions, or an intermediate configuration~\cite{ALICE:2024whv}. This uncertainty motivates a comparison of the \NplusN{} and \NplusS{} prescriptions. Here, $N$ and $S$ denote a full nucleus and a spectator target, respectively; the first and second symbols specify the photon emitter and photonuclear target~\cite{Zha:2018}. Their different overlap contributions affect both the coherent yield and the dilepton coefficient $\Atwo$, allowing these observables to constrain the production geometry jointly.

In this work, we calculate the dilepton coefficient $\Atwo$ relative to the reaction plane for coherent $J/\psi$ photoproduction in hadronic Pb--Pb collisions at $\snn=5.36$~TeV. We use an equivalent-photon and generalized VMD--Glauber framework with finite longitudinal coherence and two-source interference, and compare \NplusN{} and \NplusS{} across transverse momentum, rapidity, and impact parameter. The calculation covers the production stage and excludes subsequent in-medium $J/\psi$ dissociation and regeneration.

\section{Calculation framework}

We take the beam direction as the $z$ axis and choose the $x$ axis parallel to the impact-parameter vector, so that $\Phirp=0$. Quasireal photons are linearly polarized along the local transverse electric field of the emitting nucleus~\cite{Li:2019}. Under the helicity no-flip assumption, this polarization is transferred to the produced vector meson~\cite{Schilling:1970,Mathieu:2018}. In the ideal limit of negligible source extent, the polarization aligns with the impact-parameter direction. We denote the polarization projection along this direction by $P_{\gamma}$ and write the dilepton decay angular distribution as~\cite{Faccioli:2010}
\begin{align}
\frac{d^{2}N}{d\cos\theta\,d\phi}
=&\frac{3}{16\pi}(1+\cos^{2}\theta)
\nonumber\\
&\times\left[1-P_{\gamma}
\frac{\sin^{2}\theta}{1+\cos^{2}\theta}
\cos2(\phi-\Phirp)\right],
\label{eq:decaypgamma}
\end{align}
where $\theta$ and $\phi$ are the polar and azimuthal angles of a daughter lepton in the vector-meson rest frame, and complete alignment corresponds to $P_{\gamma}=1$. We quantify the second-order modulation by
\begin{equation}
\Atwo
=
\left\langle\cos 2(\phi-\Phirp)\right\rangle
=
\left\langle\cos 2\phi\right\rangle.
\label{eq:a2def}
\end{equation}
For full decay-lepton acceptance, integration over the polar angle gives
\begin{equation}
\frac{dN}{d\phi}=\frac{1}{2\pi}
\left[1-\frac{P_{\gamma}}{2}\cos2(\phi-\Phirp)\right],
\label{eq:azimuthpgamma}
\end{equation}
and hence $\Atwo=-1/4$ for complete polarization alignment.

For finite nuclei, the locally radial photon-polarization direction varies across the production region. We therefore retain these local directions when coherently summing the production amplitudes across the two nuclei to determine the yield and polarization projection. At a transverse production point $\xperp$, $\rvec_i$ denotes its displacement from the center of nucleus $i$, with $r_i=|\rvec_i|$ and local polarization direction $\hat{\bm e}_i=\rvec_i/r_i$. The coordinate-space production amplitude is
\begin{align}
\bm{A}_i(\xperp;\bvec,y)=&\;\hat{\bm e}_i
\left[\omega_{\gamma,i} n(\omega_{\gamma,i},r_i)\right]^{1/2}
\nonumber\\
&\times\Gamma_{\gamma A}(\omega_{\gamma,i},r_{\bar i})
S_i(\xperp;\bvec),
\label{eq:coordamp}
\end{align}
where $i$ labels the photon emitter and $\bar i$ the target nucleus. Here $n$ is the photon number density per unit energy and transverse area, $\Gamma_{\gamma A}$ describes coherent photonuclear scattering, and $S_i$ weights the target contribution in the overlap region. For a vector meson of mass $M_V$ and rapidity $y$, the two photon energies are
\begin{equation}
\omega_{\gamma,i}=\frac{M_V}{2}e^{\pm y}.
\label{eq:omega}
\end{equation}
The photon flux is evaluated in the equivalent-photon approximation. We use the finite nuclear charge form factor at transverse distances below 10 fm and the point-charge limit at larger distances. The finite-size expression is~\cite{Klein:1999qj,Baltz:2008}
\begin{equation}
\begin{aligned}
 n(\omega,r)=&\frac{4Z^{2}\alpha}{\omega}
 \left|\int\frac{d^{2}q_{\perp}}{(2\pi)^{2}}\right.
 \\
 &\qquad\left.\times
 \frac{\bm q_{\perp}F\!\left(q_{\perp}^{2}+\omega^{2}/\gamma_{c}^{2}\right)}
 {q_{\perp}^{2}+\omega^{2}/\gamma_{c}^{2}}
 e^{i\bm q_{\perp}\cdot\bm r}\right|^{2},
\end{aligned}
\label{eq:flux}
\end{equation}
where $\omega$ is the photon energy, $F$ is the nuclear charge form factor, and $\gamma_c$ is the beam Lorentz factor. A hard-sphere--Yukawa parametrization approximating a Woods--Saxon charge distribution is used~\cite{Klein:1999qj}.

The nuclear geometry entering the scattering profile is specified by the thickness function
\begin{equation}
T_A(r_{\perp})=\int_{-\infty}^{+\infty}dz\,
\rho_A\!\left(\sqrt{r_{\perp}^{2}+z^{2}}\right),
\label{eq:thickness}
\end{equation}
where $\rho_A(r)$ is the nucleon density. We use the Woods--Saxon form
\begin{equation}
\rho_A(r)=\frac{\rho_0}{1+\exp[(r-R_{\rm WS})/d]},
\label{eq:ws}
\end{equation}
with nuclear radius $R_{\rm WS}=6.62$~fm and surface diffuseness $d=0.546$~fm for Pb~\cite{DeVries:1987}.

Coherent scattering from the target is described by the generalized VMD--Glauber profile~\cite{Bauer:1977iq,Zha:2019slit}
\begin{equation}
\Gamma_{\gamma A}(\omega,r_{\perp})
=C\sqrt{\frac{\alpha}{F_V}}
\left[1-e^{-\frac{1}{2}\sigma_{VN}(W)T'_A(q_L,r_{\perp})}\right],
\label{eq:glauberprofile}
\end{equation}
where $\sigma_{VN}$ is the effective $J/\psi$--nucleon total cross section, $F_V=f_V^{2}/(4\pi)=10.4$ is the vector-meson--photon coupling~\cite{Klein:1999qj}, and $C=0.30$ accounts for non-diagonal VMD transitions~\cite{Hufner:1997jg}. Finite longitudinal coherence is included through the phase-weighted thickness~\cite{Zha:2019slit}
\begin{equation}
T'_A(q_L,r_{\perp})=
\int_{-\infty}^{+\infty}dz\,
\rho_A\!\left(\sqrt{r_{\perp}^{2}+z^{2}}\right)e^{iq_Lz},
\label{eq:phaseweightedthickness}
\end{equation}
with longitudinal momentum transfer $q_L=M_V^{2}/(4\gamma_c\omega_{\gamma})$.

To determine $\sigma_{VN}$, we use the elementary $\gamma p\rightarrow J/\psi p$ cross section fitted to world data~\cite{Cao:2018hvy},
\begin{equation}
\sigma_{\gamma p}(W)
=
C_0
\left[
1-\frac{(m_p+M_{J/\psi})^2}{W^2}
\right]^{3/2}
\left(
\frac{W^2}{W_0^2}
\right)^\delta,
\label{eq:gammap}
\end{equation}
where $C_0=80.2$~nb, $\delta=0.321$, and $W_0=100$~GeV. The photon--proton center-of-mass energy $W$ is evaluated from the photon energy in the collider frame. The forward differential cross section is obtained with a constant exponential slope parameter~\cite{Klein:1999qj},
\begin{equation}
\left.
\frac{d\sigma_{\gamma p}}{dt}
\right|_{t=0}
=
B_{J/\psi}\,\sigma_{\gamma p}(W),
\qquad
B_{J/\psi}=4.0~\mathrm{GeV}^{-2},
\label{eq:forwardgammap}
\end{equation}
and generalized VMD together with the optical theorem gives
\begin{equation}
\sigma_{VN}(W)=\frac{1}{C}
\left[
\frac{16\pi F_V}{\alpha}
\left.\frac{d\sigma_{\gamma p}}{dt}\right|_{t=0}
\right]^{1/2}.
\label{eq:sigmavn}
\end{equation}

The overlap weight $S_i$ in Eq.~\eqref{eq:coordamp} is specified by the \NplusN{} and \NplusS{} prescriptions~\cite{Zha:2018}. Photon emission is evaluated over the full nucleus in both cases. In \NplusN{}, coherent scattering extends over the full target profile, with $S_i=1$. In \NplusS{}, the amplitude is weighted by the probability that a target nucleon undergoes no inelastic interaction with the opposing nucleus~\cite{Wu:2022},
\begin{equation}
P_{\rm noH}^{(i)}(\xperp;\bvec)=
\left[1-\sigma_{NN}\frac{T_i(r_i)}{A_i}\right]^{A_i},
\label{eq:pnoh}
\end{equation}
where $T_i$ and $A_i$ are the thickness function and mass number of the opposing nucleus. We use $\sigma_{NN}=6.8$~fm$^{2}$~\cite{Loizides:2017ack}. Setting $S_i=P_{\rm noH}^{(i)}$ continuously suppresses the target's overlap contribution.

Because the two emitter--target processes are indistinguishable, their coordinate-space amplitudes are summed before Fourier transformation to momentum space,
\begin{align}
\bm{A}(\pperp;\bvec,y)=&\frac{1}{2\pi}
\int d^{2}x_{\perp}\,e^{i\pperp\cdot\xperp}
\nonumber\\
&\times\left[\bm{A}_1(\xperp;\bvec,y)+\bm{A}_2(\xperp;\bvec,y)\right].
\label{eq:fourier}
\end{align}
The coherent differential cross section is
\begin{equation}
\frac{d\sigma}{dy\,d^{2}p_{\perp}\,d^{2}b}
=|A_x|^{2}+|A_y|^{2}\equiv I.
\label{eq:coherentcross}
\end{equation}
The no-interference reference omits cross terms between the two emitter--target amplitudes while retaining the spatial coherence of each source.

The relative strengths and phases of the transverse components $\bm{A}=(A_x,A_y)$ also determine the polarization entering the decay distribution. At nonzero rapidity, these components can acquire a relative phase, allowing the net polarization to become elliptical. Following Ref.~\cite{Wu:2022}, the effective projection along the impact-parameter direction is
\begin{equation}
P_{\gamma}=
\frac{|A_x|^{2}-|A_y|^{2}}
{\sqrt{I^{2}-4\left[\operatorname{Im}\!\left(A_x A_y^{*}\right)\right]^{2}}}.
\label{eq:pgamma}
\end{equation}
At midrapidity, where the two transverse components have a common phase, this expression reduces to
\begin{equation}
P_{\gamma}=\frac{|A_x|^{2}-|A_y|^{2}}
{|A_x|^{2}+|A_y|^{2}}.
\label{eq:pgammamid}
\end{equation}
Using this projection in Eq.~\eqref{eq:azimuthpgamma} and averaging over the selected phase space gives
\begin{equation}
\Atwo=-\frac14\langle P_{\gamma}\rangle_I,
\label{eq:a2pgamma}
\end{equation}
where the average is weighted by the coherent intensity $I$. A purely $x$-polarized amplitude ($A_y=0$) gives $P_{\gamma}=1$, recovering the ideal value $\Atwo=-1/4$ discussed above. Departures from this limit reflect the local polarization directions and their coherent superposition. Applying the same yield-weighted average at fixed $\pt$ gives
\begin{equation}
\Atwo(\pt)=-\frac14
\frac{\displaystyle\iiint 2\pi b\,I P_{\gamma}\,dy\,db\,d\phi_p}
{\displaystyle\iiint 2\pi b\,I\,dy\,db\,d\phi_p}.
\label{eq:a2pt}
\end{equation}
Here $\phi_p$ is the azimuthal angle of the vector-meson transverse momentum. For $\Atwo(y)$ and $\Atwo(b)$, the same yield-weighted average is evaluated at fixed $y$ and fixed $b$, respectively, with the other variables integrated over the ranges specified in the figure captions.

\section{Results and discussion}

\begin{figure}[t]
\centering
\includegraphics[width=0.96\columnwidth]{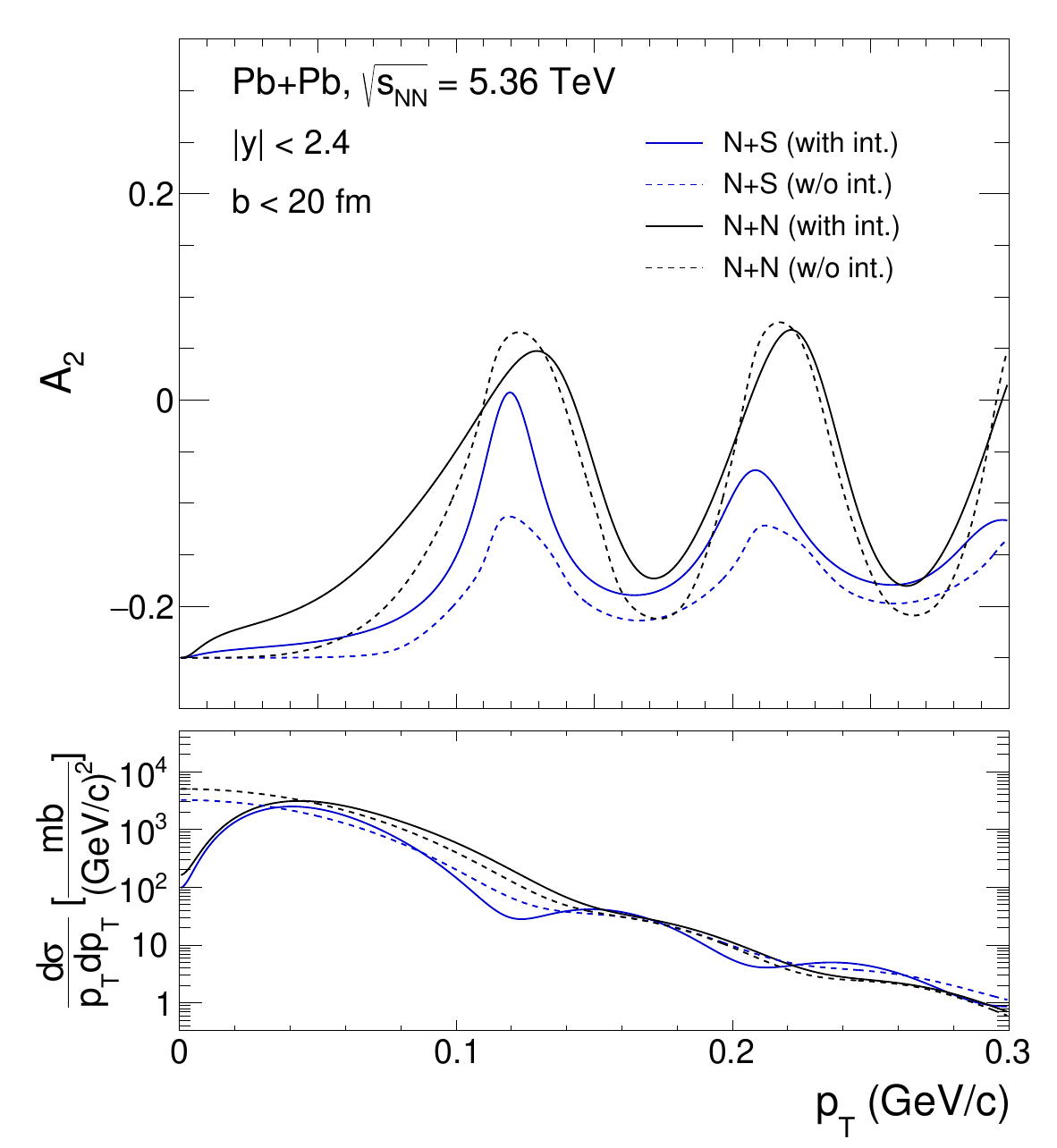}
\caption{
Upper panel: dilepton coefficient $\Atwo$ as a function of the $J/\psi$ transverse momentum for $|y|<2.4$ and $b<20$~fm in Pb--Pb collisions at $\snn=5.36$~TeV.
Lower panel: corresponding coherent production spectrum $d\sigma/(\pt d\pt)$.
The black curves show the \NplusN{} calculation and the blue curves show the \NplusS{} calculation. Solid curves include two-source interference, while dashed curves show the corresponding no-interference references.
}
\label{fig:pt}
\end{figure}

\begin{figure}[t]
\centering
\includegraphics[width=0.96\columnwidth]{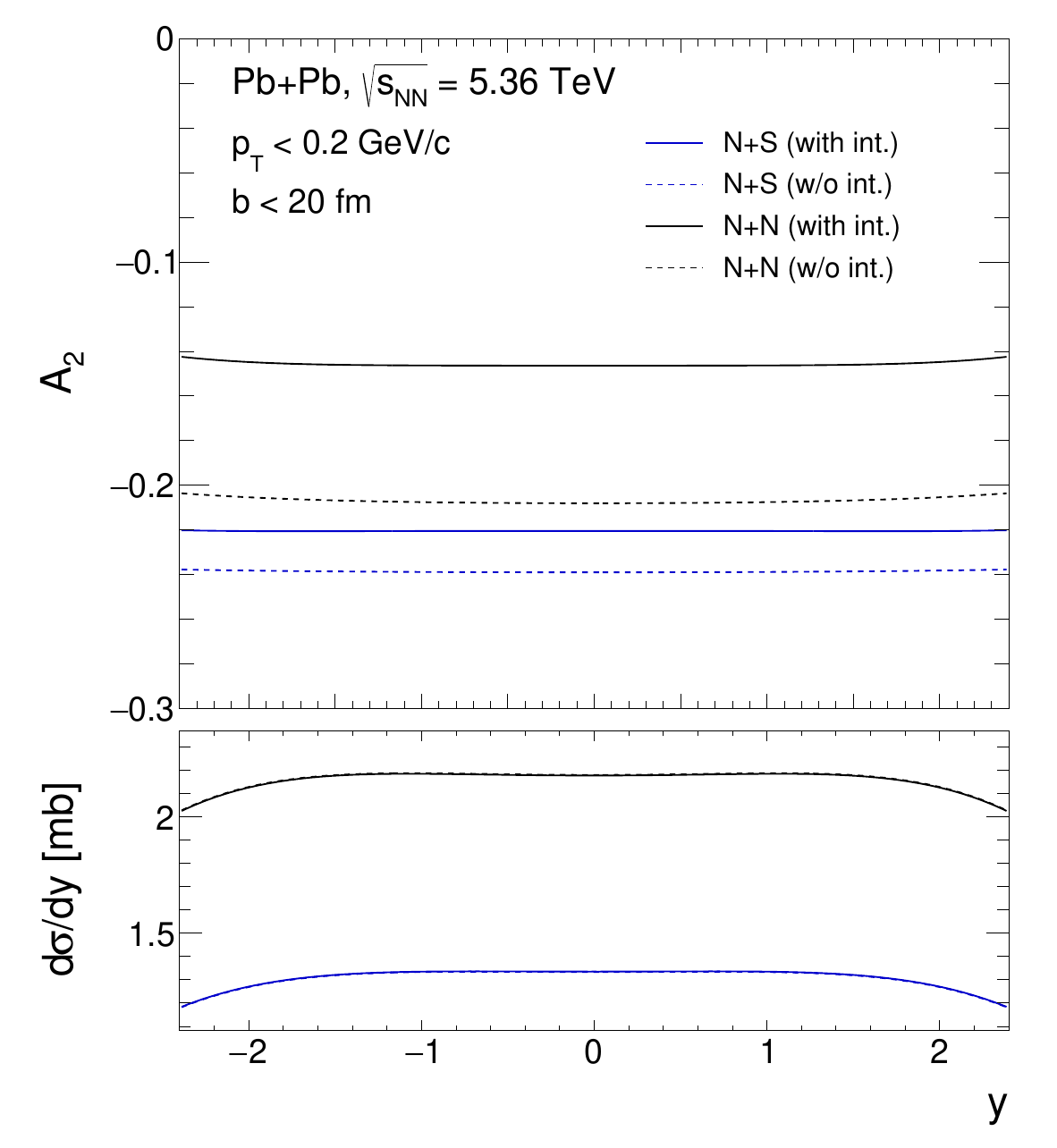}
\caption{
Upper panel: $\Atwo$ as a function of the $J/\psi$ rapidity for $\pt<0.2$~GeV/$c$ and $b<20$~fm.
Lower panel: corresponding coherent rapidity distribution $d\sigma/dy$.
The black curves show the \NplusN{} calculation and the blue curves show the \NplusS{} calculation. Solid curves include two-source interference, while dashed curves show the corresponding no-interference references.
}
\label{fig:y}
\end{figure}

\begin{figure}[t]
\centering
\includegraphics[width=0.96\columnwidth]{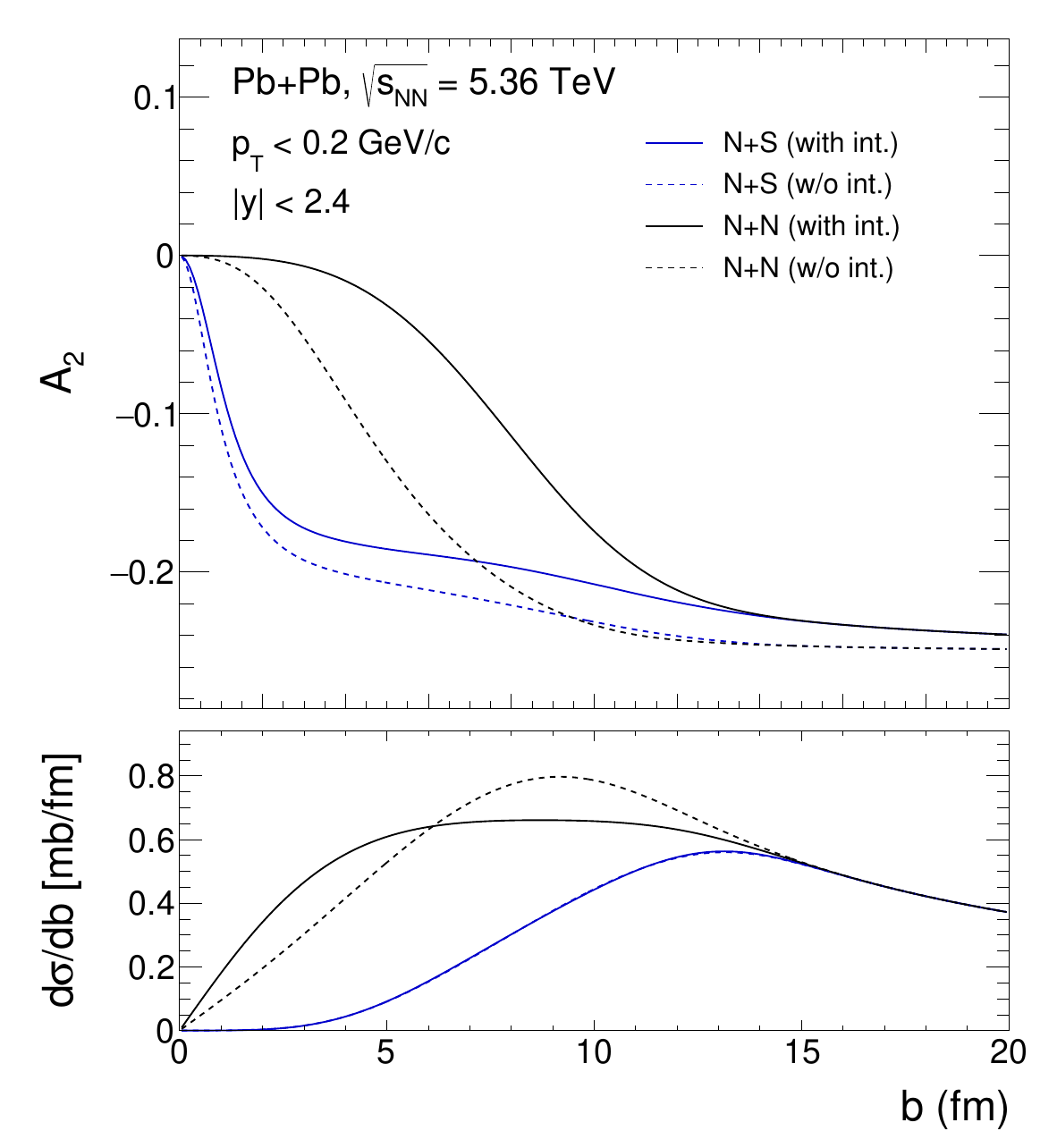}
\caption{
Upper panel: $\Atwo$ as a function of the true impact parameter for $|y|<2.4$ and $\pt<0.2$~GeV/$c$.
Lower panel: corresponding impact-parameter distribution $d\sigma/db$.
The black curves show the \NplusN{} calculation and the blue curves show the \NplusS{} calculation. Solid curves include two-source interference, while dashed curves show the corresponding no-interference references.
}
\label{fig:b}
\end{figure}

Figures~\ref{fig:pt}--\ref{fig:b} compare the coherent production distributions and reaction-plane-dependent dilepton coefficient $\Atwo$ for \NplusN{} and \NplusS{} in Pb--Pb collisions at $\snn=5.36$~TeV. Solid curves include two-source interference, while dashed curves show the corresponding no-interference references. The yield and $\Atwo$ probe the strength and polarization structure of the same production amplitude.

Figure~\ref{fig:pt} shows oscillatory structures in both the coherent production spectrum and $\Atwo(\pt)$, reflecting diffraction from the finite transverse extent of the coherent source. Two-source interference further modifies the positions and depths of the peaks and valleys. For the negative-parity $J/\psi$, destructive interference suppresses the spectrum toward $\pt=0$, with complete cancellation for equal source amplitudes, as at midrapidity in symmetric collisions~\cite{Klein:2000}. The relative two-source phase varies across momentum space, producing alternating destructive and constructive contributions. These largely compensate after integration with the $\pt\,d\pt$ measure, leaving the yield over the displayed range nearly unchanged.

In the long-wavelength limit $\pt\to0$, $\Atwo$ approaches $-1/4$, indicating nearly complete polarization alignment with the impact-parameter axis, as in large-$b$ ultraperipheral configurations. At finite $\pt$, the separation of the curves reflects the effective coherent production profiles in \NplusN{} and \NplusS{}, together with two-source interference. Suppressing the overlap contribution in \NplusS{} reduces the yield and downweights contributions with a broader spread of local polarization directions, strengthening the reaction-plane correlation. Interference modifies the vector structure of this source through coherent addition of the two amplitudes. In parts of the overlap region, the components parallel to $\bvec$ can partially cancel, while the perpendicular components can add. These opposite corrections can largely cancel in the unpolarized intensity but remain substantial in the polarization difference, reducing $|\Atwo|$ over broad momentum intervals even when the integrated yield changes little. The transverse components also weight the spatial source differently, so their diffraction minima need not coincide. Changes in their relative magnitudes and phases can produce rapid variations of $\Atwo(\pt)$, whose extrema need not follow those of the unpolarized spectrum. Modifying the source profile shifts both the diffractive patterns and their relative polarization weights, allowing local crossings between the curves. A large variation of the normalized coefficient near a spectral minimum can therefore coexist with a small production yield.

Changing rapidity varies the photon energies and the relative strengths of the two emitter--target configurations. After integration over transverse momentum, Fig.~\ref{fig:y} shows a nearly rapidity-independent negative $\Atwo$, whereas $d\sigma/dy$ has a broad maximum near midrapidity and decreases toward larger $|y|$. At LHC energies, the large beam Lorentz factor keeps $\omega/\gamma_c$ in Eq.~\eqref{eq:flux} small on the nuclear transverse-momentum scale, making the transverse photon-field profile only weakly dependent on rapidity. Longitudinal dephasing remains weak: $q_LR_{\rm WS}\lesssim0.2$ over the studied rapidity range, so $T'_A$ in Eq.~\eqref{eq:phaseweightedthickness} stays close to $T_A$ and longitudinal nuclear coherence is largely maintained. The yield retains the energy dependence of the photon flux and photonuclear scattering strength; its symmetry under $y\to-y$ follows from exchanging the two identical nuclei. The \NplusS{} predictions remain lower in yield and larger in $|\Atwo|$. At midrapidity, interference shifts $\Atwo$ from $-0.208$ to $-0.146$ for \NplusN{} and from $-0.239$ to $-0.221$ for \NplusS{}, reducing their magnitudes by approximately $30\%$ and $8\%$, respectively. These shifts vary little over the studied rapidity range. The difference between the interference shifts in \NplusN{} and \NplusS{} originates from the overlap contribution: as discussed above, interference can weaken the polarization alignment with $\bvec$ through partial cancellation of the parallel components. Attenuating this contribution in \NplusS{} therefore reduces the interference correction relative to \NplusN{}.

Figure~\ref{fig:b} resolves the collision geometries averaged over in the transverse-momentum and rapidity distributions. In the lower panel, $d\sigma/db$ vanishes as $b\to0$ owing to the geometrical factor $2\pi b$. Two-source interference redistributes the coherent yield over impact parameter: for \NplusN{}, it enhances $d\sigma/db$ at small $b$ and suppresses it over part of the intermediate-$b$ region. These opposite corrections largely compensate after integration over $b$, leaving the total yield within the selected kinematic and impact-parameter ranges nearly unchanged. In \NplusS{}, attenuating the overlap contribution reduces the yield at small and intermediate $b$ and weakens the spatial overlap of the two production amplitudes. Interference consequently has a much smaller effect on the shape of the momentum-integrated $b$ distribution.

In the upper panel, all four calculations give $\Atwo\to0$ as $b\to0$, reflecting the absence of a preferred transverse direction in the smooth nuclear geometry; the reaction plane is undefined at exactly $b=0$. At nonzero $b$, attenuation of the overlap contributions in \NplusS{} strengthens the polarization alignment with the impact-parameter axis and reduces its sensitivity to interference. \NplusN{} retains these contributions and therefore exhibits a more pronounced reduction of $|\Atwo|$ from interference at intermediate $b$. As $b$ increases further, the shrinking overlap brings the two prescriptions closer, while the photon linear-polarization axes across each target become increasingly aligned with the impact-parameter axis. For well-separated nuclei, the geometric angular spread is of order $R_{\rm WS}/b$, so all four calculations approach the ideal limit $\Atwo=-1/4$, with finite-size corrections remaining at the impact parameters shown. After integration over $b$, the suppression of small-$b$ production in \NplusS{} further strengthens the correlation by increasing the relative weight of peripheral configurations with larger $|\Atwo|$.

The different sensitivities of the coherent yield and $\Atwo$ to the overlap contribution motivate joint measurements to constrain the production geometry. Measurements of very-low-$\pt$ $J/\psi$ production show that nuclear coherence can survive in collisions with hadronic overlap~\cite{STAR:2019yox}. Yield comparisons remain compatible, within uncertainties, with several treatments of the overlap region~\cite{ALICE:2024whv}. Polarization adds a directional constraint: sources with similar integrated strengths can differ in how those strengths are distributed between components parallel and perpendicular to the impact-parameter axis.

Decay-anisotropy measurements by STAR at RHIC and CMS at the LHC provide access to this directional information. STAR reports a photon-induced $J/\psi$ decay anisotropy relative to the hadronic event plane in isobar collisions~\cite{Wang:2026STAR}. CMS observes a negative coefficient in peripheral Pb--Pb collisions at $\snn=5.36$~TeV, with no significant centrality or rapidity dependence within the present uncertainties~\cite{CMS:2026}. A color-dipole calculation describes the STAR trend~\cite{Shao:2025}, although its LHC prediction exceeds the measured magnitude according to the CMS analysis~\cite{CMS:2026}.

As a geometrical estimate, the CMS 50--60\% centrality interval corresponds approximately to $b\simeq11.0$--$12.1$~fm in Pb--Pb Glauber calculations at nearby LHC energies~\cite{Loizides:2017ack}. In Fig.~\ref{fig:b}, which integrates over $|y|<2.4$, the curves including interference vary across this impact-parameter interval by about $0.016$ for \NplusN{} and $0.006$ for \NplusS{}, while the separation between the prescriptions is already reduced. The CMS interval therefore lies in a peripheral region where the predicted sensitivity to the production geometry is weaker. Both prescriptions also predict a weak rapidity dependence, so the absence of a resolved variation in the CMS data does not by itself distinguish between them.

A quantitative comparison requires integrating the polarization-weighted numerator and coherent yield over the CMS selections, $1.6<|y|<2.4$ and $\pt<0.2$~GeV/$c$, and over the impact-parameter range corresponding to each experimental centrality class, using $2\pi b\,db$. The decay-angle definitions and the relation between the reconstructed event plane and the reaction plane must also be treated consistently~\cite{CMS:2026}. When the measured coefficient is defined within finite decay-lepton acceptance, the polar-angle integration in Eq.~\eqref{eq:decaypgamma} must also be restricted accordingly. For example, for a midrapidity $J/\psi$ with negligible $\pt$, a central selection $|\eta_{\ell}|<0.9$ favors decays near $\theta=\pi/2$, where the factor $\sin^{2}\theta/(1+\cos^{2}\theta)$ is larger. For $P_{\gamma}=1$ and uniform azimuthal acceptance, this selection changes the ideal coefficient from $-1/4$ to approximately $-0.354$, an increase of about $42\%$ in magnitude. A comparison with a fiducial measurement therefore requires applying the same lepton acceptance to the prediction.

Larger data samples would enable finer centrality and rapidity measurements and extend the reach toward more central collisions, where the predicted separation between \NplusN{} and \NplusS{} is larger. Combined with coherent-yield measurements and theoretical uncertainty estimates, these measurements can help distinguish the two prescriptions and determine how strongly the overlap region participates in coherent production.

The polarization--geometry connection also motivates an extension to small systems, where nonflow correlations complicate studies of collectivity~\cite{GrosseOetringhaus:2024}. With sufficient reconstructed photoproduced $J/\psi$ mesons, their decay anisotropy could provide an independent statistical reference for the impact-parameter direction, enabling correlations between decay leptons and soft particles to test the geometrical origin of soft-particle anisotropy~\cite{Wu:2022}. Geometry fluctuations can, however, weaken the alignment between the participant plane and the impact-parameter direction~\cite{Alver:2010gr}. The sensitivity of this test therefore depends on both the tagged yield and that geometrical correlation. Dedicated calculations of the polarized yield and fluctuating geometry in light-ion, $pA$, and $pp$ collisions are needed to assess the feasibility of extending this approach to small systems. For light-ion systems, existing calculations of O--O ultraperipheral collisions have already explored the effects of gluon-field fluctuations and $\alpha$-cluster structure on coherent and incoherent vector-meson photoproduction cross sections~\cite{Cepila:2026OO}. The present calculation uses smooth nuclear densities without explicit event-by-event configurations. Comparing future measurements of the coherent yield and $\Atwo$ in light-ion collisions with predictions based on smooth densities and on configurations incorporating nuclear structure and fluctuations could test the applicability of this smooth-density description.

\section{Summary}

We have studied the correlation between coherent $J/\psi$ photoproduction and the reaction plane in hadronic Pb--Pb collisions at $\snn=5.36$~TeV. The calculation uses the equivalent-photon and generalized VMD--Glauber framework, including finite longitudinal coherence and two-source interference, to compare the \NplusN{} and \NplusS{} prescriptions.

Suppressing the overlap contribution in \NplusS{} reduces the coherent yield and increases the magnitude of the integrated $\Atwo$. The difference in $\Atwo(b)$ between the prescriptions is largest at small but nonzero and intermediate impact parameters and decreases toward peripheral collisions. All four calculations approach the ideal limit $\Atwo=-1/4$ at large $b$. Nuclear diffraction and interference shape the transverse-momentum dependence of the yield and $\Atwo$, while $\Atwo$ varies weakly with rapidity. Interference changes the polarization correlation more strongly for \NplusN{} than for \NplusS{}, while leaving the yield integrated over the studied phase space nearly unchanged.

Joint measurements of the coherent yield and $\Atwo$ can therefore test how the overlap region participates in coherent production. The STAR and CMS decay-anisotropy measurements motivate further tests with larger data samples and finer centrality selections. With sufficient tagged yields and a measurable correlation between the participant plane and the impact-parameter direction, the decay anisotropy could provide an independent reference for studies of collectivity. Future measurements in light-ion collisions, compared with calculations incorporating nuclear structure and fluctuations, could also test the applicability of the smooth-density description.

\section*{Acknowledgments}

This work was supported in part by the National Natural Science Foundation of China under Grant Nos.~12605233, 12422510, 12361141827, and 12275091, and the National Key Research and Development Program of China under Grant No.~2022YFA1604900. W.~Zha also acknowledges support from the Anhui Provincial Natural Science Foundation under Grant No.~2508085JX002, the Youth Innovation Promotion Association of the Chinese Academy of Sciences, and the Chinese Academy of Sciences under Grant No.~YSBR088.

\bibliographystyle{nst_unsrt}
\bibliography{references}

\end{document}